\documentclass[aps,prb,superscriptaddress,floatfix,twocolumn,longbibliography]{revtex4-2}

\usepackage{amsmath,dcolumn,bm,amsthm,tabularx,subfigure}
\usepackage[colorlinks=true,urlcolor=blue,citecolor=blue,linkcolor=blue]{hyperref}
\usepackage{mathrsfs,times}
\usepackage{bbold,dsfont}
\usepackage{graphicx,graphics,color}
\usepackage{mathtools,nicefrac}
\usepackage{slashed}
\usepackage{amsfonts,amssymb}
\usepackage{multirow}
\usepackage{booktabs}
\allowdisplaybreaks

\newcommand{\beq}{\begin{equation}}
\newcommand{\eeq}{\end{equation}}
\newcommand{\bea}{\begin{eqnarray}}
\newcommand{\eea}{\end{eqnarray}}

\begin{document}

\title{Many-body localization and particle multioccupancy in the disordered Bose-Hubbard model}

\author{Jie~Chen}
\thanks{These authors contributed equally to this work}
\email[\\Contact author:\ ]{chenjie666@xhu.edu.cn}
\affiliation{Key Laboratory of Artificial Structures and Quantum Control (Ministry of Education), School of Physics and Astronomy, Shanghai Jiao Tong University, Shanghai 200240, China}
\affiliation{School of Science, Key Laboratory of High Performance Scientific Computation, Xihua University, Chengdu 610039, China}

\author{Chun~Chen}
\thanks{These authors contributed equally to this work}
\email[\\Contact author:\ ]{chunchen@sjtu.edu.cn}
\affiliation{Key Laboratory of Artificial Structures and Quantum Control (Ministry of Education), School of Physics and Astronomy, Shanghai Jiao Tong University, Shanghai 200240, China}

\author{Xiaoqun~Wang}
\email[Contact author:\ ]{xiaoqunwang@zju.edu.cn}
\affiliation{School of Physics, Zhejiang University, Hangzhou 310058, Zhejiang, China}
\affiliation{Collaborative Innovation Center of Advanced Microstructures, Nanjing University, Nanjing 210093, China}

\date{\today}

\begin{abstract}

We study the potential influence of the particle multi-occupations on the stability of many-body localization in the disordered Bose-Hubbard model. Within the higher-energy section of the dynamical phase diagram, we find that there is no apparent finite-size boundary drift between the thermal phase and the many-body localized regime. We substantiate this observation by introducing the Van Vleck perturbation theory into the field of many-body localization. The appropriateness of this method rests largely on the peculiar Hilbert-space structure enabled by the particles' Bose statistics. The situation is reversed in the lower-energy section of the dynamical phase diagram, where the significant finite-size boundary drift pushes the putative many-body localized regime up to the greater disorder strengths. We utilize the algebraic projection method to make a connection linking the disordered Bose-Hubbard model in the lower-energy section to an intricate disordered spin chain model. This issue of the finite-size drift could hence be analogous to what happens in the disordered Heisenberg chain. Both trends might be traced back to the particles' intrinsic or emergent single-occupancy constraint like the spin-$1/2$, hard-core boson, or spinless fermion degrees of freedom.  

\end{abstract}

\maketitle


\section{Introduction}

The current research on many-body localization (MBL) is severely hindered by the finite-size drift of the boundary between the thermal phase and the putative MBL regime \cite{morningstar2022avalanches,Suntajs,Sierant2024,KieferEmmanouilidisPRL,vsuntajs2020ergodicity,long2023phenomenology,aceituno2024ultraslow}. This kind of drift arises in plenty of theoretical models including the disordered Heisenberg spin chain \cite{Luitz,sierant2020polynomially,colbois2024interaction}, literally plaguing most numerical studies in the field and impeding our understanding of the MBL phenomenon.

In this regard, search for a new model system without this issue could be instrumental in the broader research endeavor to pursue the nonergodic eigenstate matter. As advocated by the present work, the disordered Bose-Hubbard (dBH) model \cite{hopjan2020many,orell2019probing,Sierant_2018,yao2020many,Filippone,Molignini_boson} realizable in cold-atom laboratories \cite{LukinGreiner,leonard2023probing,YanPRL,YanPRA,YanarXiv} might right provide such a candidate system. Concretely, stimulated by the question concerning the relationship between MBL and particle statistics in general and the clustering of interacting bosons in random potentials in particular, we identify the robust and peculiar localization signature that is suggestive of the existence of a novel cluster MBL regime in the higher-energy section of the dBH model's dynamical phase diagram. Our results hint that it is not very likely for the dBH chain to fully thermalize even at the weak disorder when the energy density is high. The claimed cluster MBL regime can thus avoid the problem of the finite-size drift and persist still in the thermodynamic limit.

To facilitate the investigation, several new methods are proposed. On the higher-energy side, we first introduce the Van Vleck perturbation theory \cite{VanVleck,CohenTannoudji,MIT} into the study of MBL, which exploits the formation of the well-separated eigenstate manifolds and the damped hybridizing between the slow intra- and the fast inter-manifold degrees of freedom given the clustering endowed by the particles' Bose statistics. Next, on the lower-energy side, we instead use the algebraic projection method \cite{Cazalilla,basak2021strongly} to expose the resembling between the dBH model and a disordered spin chain model, where the emergent spin or Fermi statistics plays a key role.

The paper is structured as follows. The investigated model is introduced in Sec.~\ref{secmodel}. The main features of the calculated dynamical phase diagrams of the model are presented and highlighted in Sec.~\ref{secphasediagram}. The peculiarity of the higher-energy section is tackled by the introduced Van Vleck perturbation theory in Sec.~\ref{secvv} where we stress the importance of the particle multi-occupancy allowed by the Bose statistics. To understand the pronounced finite-size drift seen in the lower-energy section, we invoke the algebraic projection method in Sec.~\ref{sec_fermi} to map this model to a disordered spin chain. Finally, Sec.~\ref{secdiscon} details the comparison to the earlier literature and the conclusion. 

\begin{figure*}[t]
\begin{center}
\includegraphics[width=1\linewidth]{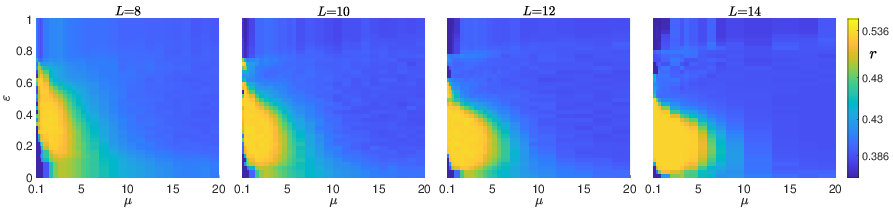}
\caption{The sequence of the small-size dynamical phase diagrams of the dBH chain model (\ref{hamdbh}) with $N=\frac{L}{2}$. The color contours are derived from the level-spacing ratio $r$ computed by ED and averaged over a sufficient amount of random samples (see the main text for details). Here, the axes $\varepsilon$ and $\mu$ stand for the normalized energy and the disorder strength, respectively.}
\label{pic8}
\end{center}
\end{figure*}  

\section{Model} \label{secmodel}

{\color{black} The dBH Hamiltonian under the periodic boundary conditions is described by}
\begin{align}
H_{\textrm{dBH}}=&-\sum^L_{i=1}J(a^{\dagger}_ia_{i+1}+\textrm{H.c.})+\sum^L_{i=1}\mu_in_i \nonumber \\
&+\sum^L_{i=1}\frac{U}{2}n_i(n_i-1)
\label{hamdbh}
\end{align}
where $a^\dagger_i\ (a_i)$ is the boson creation (annihilation) operator at site $i$, $n_i=a^\dagger_ia_i\ (N=\sum^L_in_i)$ counts the local (total) boson occupation number, $U$ parametrizes the onsite Hubbard repulsion, and $\mu_i\in[-\mu,\mu]$ is a diagonal random potential drawn from the box distribution. {\color{black} Here, the periodic boundary conditions mean that the site $i+L$ is the same as the site $i$ in the model (\ref{hamdbh}), i.e., $a_{i+L}=a_i$ and $a^\dagger_{i+L}=a^\dagger_i$ for $i=1,...,L$.} Crucially, $[N,H_{\textrm{dBH}}]=0$, so the number-conserving dBH model respects the U(1) symmetry. In this work, all the relevant quantities are the averages over a sufficient amount of the random samples, solved by exact diagonalization (ED) \cite{Zhang2010} or the Van Vleck perturbative method \cite{VanVleck}. {\color{black} Practically, it is doable that $10000$ samples are used in the averages of the data for $L=8$, $5000$ for $L=10$, $3000$ for $L=12$, and for $L=14$, $200$ samples are used for the weak disorder, $400$ for the medium disorder, and $600$ for the strong disorder.} We set $J=1$ as the energy unit and fix $U=3J,\ N=\frac{L}{2}$ in the subsequent numerical calculations.

\section{Absence of the drift in the finite-size scaling of the level-spacing ratio and the maximal site occupation} \label{secphasediagram}

First of all, it is critical practice to examine how the various phase-regime boundaries of the dynamical phase diagram of the dBH model would change under the increase of the system size. To this aim, we conduct the extensive numerical calculations to construct the finite-size dynamical phase diagrams of the dBH chain from $L=8$ up to $L=14$. Figures~\ref{pic8} and \ref{pic_ni} illustrate the obtained ED results based respectively on the measures of the level-spacing ratio \cite{Oganesyan,Atas} and the maximal site occupation. The extraction of the level-spacing ratio proceeds as follows. {\color{black} For Hamiltonian (\ref{hamdbh}) within a particular random realization, we use ED to find all its eigenvalues $\{E_n\}$ arranged in an ascending order $E_1\leqslant E_2\leqslant\cdots\leqslant E_D$ and $D$ is the dimension of the Hamiltonian matrix. The involved nearest-neighboring gap ratio is defined by
\beq
r_n=\frac{\textrm{min}(\delta E_n,\delta E_{n-1})}{\textrm{max}(\delta E_n,\delta E_{n-1})}
\eeq
where $\delta E_n=E_n-E_{n-1}$ is the gap between the two adjacent eigenenergies of $H_{\textrm{dBH}}$. The level-spacing ratio $r$ is then obtained from averaging over all the eigenenergies within an interval closest to a specified normalized energy and subsequently over all the random realizations. As in \cite{Luitz}, the normalized energy $\varepsilon$ in Fig.~\ref{pic8} is defined by $\varepsilon=(E_n-E_1)/(E_D-E_1)$.} 

There are two messages from Fig.~\ref{pic8}. First, there is a pronounced drift of the boundary between the thermal phase and the MBL regime in the lower-energy section of the dynamical phase diagram. Such a drift under the increase of $L$ toward greater disorder strengths echoes what occurs in the disordered Heisenberg chain and is known to be the obstacle toward identifying MBL as a stable phase of matter in the thermodynamic limit. Secondly, the MBL regime in the higher-energy section of the dynamical phase diagram however appears to be robust and stable. Especially, the boundary between the cluster MBL regime and the thermal phase is moving downward to the spectrum center $\varepsilon\approx0.5$, suggesting the absence of the drift and the persistence of both this MBL regime and the mobility edge in the large-size limit. Recall that at the same time, the density of states across the higher-energy section is getting enhanced upon this successive increase of $\mu$ \cite{chen2024eigenstate}. Finally, Fig.~\ref{pic_ni} shows that the two messages from Fig.~\ref{pic8} can be probed via the more accessible quantity: the maximal site occupation $\textrm{max}(n_i)/N$ as well, thereby being experimentally testable. The maximal site occupation $\textrm{max}(n_i)/N$ is calculated as follows. Once we obtain an eigenstate at a specific normalized energy, we calculate its local boson occupation number on every site of the chain. Then, by comparing and selecting the maximal expectation value of the onsite occupation number denoted as $\textrm{max}(n_i)$, the maximal site occupation $\textrm{max}(n_i)/N$ is readily calculable. The result shown in Fig.~\ref{pic_ni} is the average of $\textrm{max}(n_i)/N$ first over all the eigenstates within that normalized energy window for a random realization and then over all the available random samples.  

Hereafter, our strategy is to gain an overall understanding of the dynamical phase diagram of the dBH model by exploring the distinction in the emergent particle statistics from the opposite higher- and lower-energy limits of the phase diagram.

\section{The higher-energy limit: confirming the absence of the drift via the Van~Vleck perturbation theory} \label{secvv}

As per expression (\ref{hamdbh}), one can divide $H_{\textrm{dBH}}$ into two catergories: the off-diagonal term $H_J=-J\sum^L_{i=1}(a^{\dagger}_ia_{i+1}+\textrm{H.c.})$ and the diagonal terms $H_U=\sum^L_{i=1}\frac{U}{2}n_i(n_i-1)$ and $H_\mu=\sum^L_{i=1}\mu_in_i$. As $H_J$ is a single-particle hopping term, in the higher-energy limit, the skeleton of the $H_{\textrm{dBH}}$ matrix is built upon the $H_U$ term, which itself is structured into the individual diagonal blocks upon the basis states shared the same maximal onsite boson occupation number
\beq
n_\textrm{max}=1,\ldots,\frac{L}{2}.
\label{nmax}
\eeq
{\color{black} For instance, the basis states of the $n_\textrm{max}=3$ block are those symmetrized states in the occupation-number representation whose maximal local occupation number is $3$, meaning that there is at least one site in the chain whose occupation number is $3$, but there is no site in the chain whose occupation number is greater than $3$. Such a definition guarantees that the basis states from the different blocks do not overlap.} Clearly, $H_{\mu}$ respects this diagonal block structure. Additionally, while $H_J$ creates the off-diagonal matrix elements within each $n_\textrm{max}$ block, it also engenders the off-diagonal matrix elements connecting the $n_\textrm{max}$ block to its two neighbor blocks $n_\textrm{max}\pm1$, thus furnishing a decent \emph{block-tridiagonal} structure. Note that when $n_\textrm{max}=1,\frac{L}{2}$, there is one such neighboring block. To some extent, $n_\textrm{max}$ may be regarded as an approximate quantum number once these off-diagonal inter-block submatrices are removed. As will be shown below, the Van Vleck perturbation theory \cite{VanVleck,CohenTannoudji,MIT} serves exactly this task.

\begin{figure*}[thb]
\begin{center}
\includegraphics[width=1\linewidth]{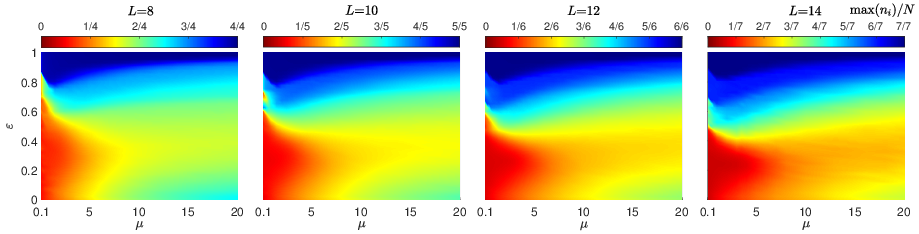}
\caption{The companion sequence of the small-size dynamical phase diagrams of the dBH chain model (\ref{hamdbh}) with $N=\frac{L}{2}$. The color contours are drawn from the quantity of the maximal site occupation $\textrm{max}(n_i)/N$ (see its definition in the main text) computed via ED and averaged over many random realizations.}
\label{pic_ni}
\end{center}
\end{figure*}

To proceed, let us temporarily neglect $H_J$ and make a comparison between the portions of the $H_U+H_\mu$ matrix within the lower- and higher-energy sections of the phase diagram, assuming that $\mu$ is small and any accidental symmetry has been broken. This diagonal $H_U+H_\mu$ matrix can be arranged into $\frac{L}{2}$ blocks according to $n_\textrm{max}$. Each eigenstate now is simply specified by the $\{n_i\}$ set; it hence represents trivially the local integrals of motion (LIOMs) \cite{Serbyn,HuseLIOM,Anushya,geraedts2017emergent} in this case. For large-$n_\textrm{max}$ blocks, bosons of these LIOMs are more concentrated on several local sites, but for small-$n_\textrm{max}$ blocks, they spread more uniformly across the whole chain. In the absence of both $H_J$ and $H_\mu$, the bottom $n_\textrm{max}=1$ block is flat and there is a gap $U$ separating it from the $n_\textrm{max}=2$ block. This gap increases linearly with $n_\textrm{max}$. But, as the energy range of the $n_\textrm{max}=2$ block is extensive, there will be a substantial energy overlap between the $n_\textrm{max}=2$ and $3$ blocks. Such a trend continues in the lower-energy section of the phase diagram. In contrast, near the top $n_\textrm{max}=\frac{L}{2}$ block, the situation alters sharply. Due to the Bose statistics, $n_\textrm{max}$ in this circumstance can be huge such that the large-$n_\textrm{max}$ blocks comprising the higher-energy section of the phase diagram are well separated in energy scale even for weak or moderate $U$: they form the distinguishing manifolds without the energy overlaps. Transparently, this occupation-number driven energy separation differs drastically from the situation of the more familiar large or infinite-$U$ limit in the fermion or spin systems, where one instead needs to do the ground-state manifold projection before taking the infinite-temperature limit. {\color{black} See below for more elaboration on this and the related points.}

The Van Vleck approximation is to devise a suitable canonical transformation $e^{iS}$ to perturbatively achieve the goal that the transformed Hamiltonian $H'_{\textrm{dBH}}=e^{iS}H_{\textrm{dBH}}e^{-iS}$ preserves the same eigenenergies with the same degeneracy as the original Hamiltonian $H_{\textrm{dBH}}$ but simultaneously has no matrix elements between the unperturbed blocks up to the desired order of the small perturbation. Focus on the higher-energy section of the phase diagram, one can choose the unperturbed Hamiltonian $H^0=H_U+H_\mu$ and the small perturbation $H_J$. By demanding the canonical transformation $S=S^\dagger$ to be fully off-block-diagonal, i.e., $\langle i,\alpha|S|j,\alpha\rangle=0$ (see definitions below), one can explicitly construct, up to third order of $J$, the transformed block diagonal Hamiltonian as follows,
\begin{widetext}
\begin{align}
\langle i,\alpha|&H'_{\textrm{dBH}}|j,\alpha\rangle=E^0_{i\alpha}\delta_{ij}+\langle i,\alpha|H_J|j,\alpha\rangle+\frac{1}{2}\sum_{k,\gamma\neq\alpha}\langle i,\alpha|H_J|k,\gamma\rangle\langle k,\gamma|H_J|j,\alpha\rangle\left(\frac{1}{E^0_{i\alpha}-E^0_{k\gamma}}+\frac{1}{E^0_{j\alpha}-E^0_{k\gamma}}\right) \nonumber \\
&+\frac{1}{2}\sum_{l,\eta\neq\alpha}\left\{\sum_{k,\gamma\neq\alpha}\left[\frac{\langle i,\alpha|H_J|k,\gamma\rangle\langle k,\gamma|H_J|l,\eta\rangle\langle l,\eta|H_J|j,\alpha\rangle}{(E^0_{i\alpha}-E^0_{k\gamma})(E^0_{i\alpha}-E^0_{l\eta})}+\frac{\langle i,\alpha|H_J|l,\eta\rangle\langle l,\eta|H_J|k,\gamma\rangle\langle k,\gamma|H_J|j,\alpha\rangle}{(E^0_{j\alpha}-E^0_{k\gamma})(E^0_{j\alpha}-E^0_{l\eta})}\right]\right\} \nonumber \\
&-\frac{1}{2}\sum_{l,\eta\neq\alpha}\left\{\sum_{k,\gamma\neq\eta}\left[\frac{\langle i,\alpha|H_J|k,\gamma\rangle\langle k,\gamma|H_J|l,\eta\rangle\langle l,\eta|H_J|j,\alpha\rangle}{(E^0_{i\alpha}-E^0_{l\eta})(E^0_{k\gamma}-E^0_{l\eta})}+\frac{\langle i,\alpha|H_J|l,\eta\rangle\langle l,\eta|H_J|k,\gamma\rangle\langle k,\gamma|H_J|j,\alpha\rangle}{(E^0_{j\alpha}-E^0_{l\eta})(E^0_{k\gamma}-E^0_{l\eta})}\right]\right\}+\cdots.
\label{Hdbhprime}
\end{align}
\end{widetext}
Here $\alpha,\gamma,\eta$ are the unperturbed block-$n_\textrm{max}$ indices. For a particular block, the involved ket and bra vectors indexed by $i,j,k,l$ are its corresponding eigenstates of the unperturbed $H^0$, i.e., $H^0|i,\alpha\rangle=E^0_{i\alpha}|i,\alpha\rangle$. Once $H'_{\textrm{dBH}}$ is available for block $\alpha$, one can exact diagonalize the resulting matrix (\ref{Hdbhprime}) to get the perturbative estimates of the eigenenergies of $H_{\textrm{dBH}}$ with respect to this emergent quantum number $n_\textrm{max}$. It is crucial that while the inter-block off-diagonal matrix elements are only eliminated perturbatively, the intra-block off-diagonal matrix elements are treated almost exactly. The obtained eigenfunctions of $H'_{\textrm{dBH}}$ would probably have degraded accuracy, so they should be used with caution to extract the wavefunction-related information, such as the entanglement entropy. 

\begin{figure}[t]
\begin{center}
\includegraphics[width=0.85\linewidth]{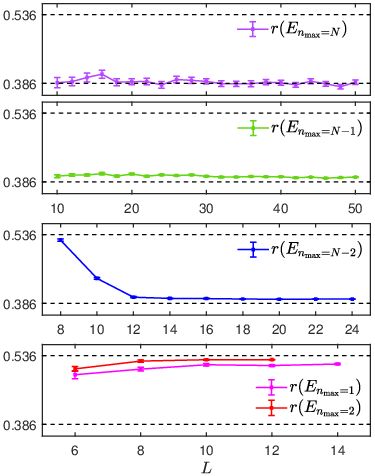}
\caption{Perturbative estimates of the level-spacing ratio $r$ for the top blocks $n_{\textrm{max}}=N,\ N-1,\ N-2$ and bottom blocks $n_{\textrm{max}}=1,\ 2$ as a function of the dBH chain length $L$. These scaling results are obtained by the Van Vleck algorithm and averaged over at least $100$ random samples at weak disorder $\mu=2J$. {\color{black} Here, the top dashed line gives the prediction of the level-spacing ratio suitable for the Gaussian orthogonal ensemble, $r_{\rm GOE}\approx0.536$, while the bottom dashed line gives the corresponding prediction for the ensemble obeying the Poisson statistics, $r_{\rm Poi.}\approx0.386$.}}
\label{pic_vv}
\end{center}
\end{figure} 

For large-$n_{\textrm{max}}$ blocks in the higher-energy section of the phase diagram, both the large intra-block $H_U$ values and the ensuring large inter-block energy gaps guarantee the comparative smallness of the perturbation $H_J$, which, in turn, renders the outlined computational scheme founded upon the Van Vleck degenerate perturbation theory particularly appealing in helping understand the cluster MBL regime at weak disorder.

{\color{black} To ensure the applicability of the Van Vleck theory, we want the energy gap between the neighboring eigenstate manifolds to be generically large or at least these manifolds are well separated in energy scale. If the disorder strength $\mu$ is too large, then the resultant broadening of the manifold will be significantly enhanced. This might cause the energy overlaps between the adjacent manifolds, formally impacting the accuracy of the perturbative calculations. Approximately, the accuracy of the Van Vleck theory is set by $(J/\Delta)^3$ where $\Delta$ quantifies the mean energy gap between the neighboring manifolds involved. Thus, typically we want $\Delta\gg J$, meaning that the disorder strength would not mix the adjacent manifolds in energy. Moreover, the central message of this work is the possible absence of the finite-size boundary drift in the higher-energy section of the dynamical phase diagram of the dBH chain. To this end, our strategy is to carefully examine the weak-disorder regime. This is because once this weak-disorder and higher-energy regime has proven to be nonergodic, one has a good reason to believe that further increase of the disorder strength shall not turn the system to the thermal phase. This cluster MBL regime at weak disorder is where the Van Vleck degenerate perturbation theory works well.}

In view of the importance of both the identification of $n_{\textrm{max}}$ and the perturbativeness of $H_J$ in setting up the algorithm, it is intriguing that the self-consistency and the applicability of the celebrated Van Vleck perturbation theory to the dBH type model are essentially enabled by the particles' Bose statistics.

\begin{figure}[t]
\begin{center}
\includegraphics[width=0.98\linewidth]{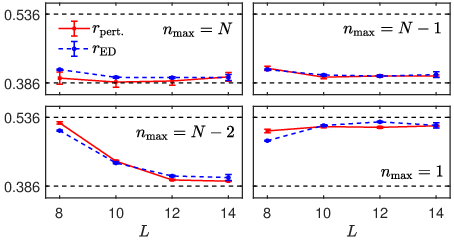}
\caption{{\color{black} Comparison between the averaged level-spacing ratio obtained from the ED, $r_{\rm ED}$, and obtained from the perturbative Van Vleck method, $r_{\rm pert.}$. The detailed procedure of how to extract $r_{\rm ED}$ from Figs.~\ref{pic8} and \ref{pic_ni} is explained in the main text. The $r_{\rm pert.}$ values are directly read out from Fig.~\ref{pic_vv}, thus here we have focused on the weak-disorder regime specified by $\mu=2J$. The two dashed lines are defined as in Fig.~\ref{pic_vv}.}}
\label{pic_pert_ed}
\end{center}
\end{figure} 

Potentially, the proposed Van Vleck algorithm could have practical and theoretical significance. On the practical side, the Hilbert-space size of the half-filled dBH chain grows rapidly with the chain length as $\frac{(3L/2-1)!}{(L/2)!(L-1)!}$. While, for $L=14$, this size is $77520$, it jumps to $490314$ for $L=16$ and $3124550$ for $L=18$. Considering the needs of random averages, this means that even for shift-invert \cite{Luitz} or polynomially filtered Lanczos \cite{sierant2020polynomially} methods, it is impractical to perform an exact scaling analysis of the pertinent quantities covering longer bosonic chains. In this regard, the Van Vleck algorithm may provide an alternative to partly tackle this problem. We explicitly test this possibility in Fig.~\ref{pic_vv} where we restrict attention to the three largest-$n_\textrm{max}$ blocks, $N$, $N-1$, and $N-2$, and succeed in executing the Van Vleck algorithm to manifest, especially from small to medium lengths, the persistent convergence of the averaged level-spacing ratio $r$ toward Poisson in a weak-disorder BH chain up to $L=50$ for blocks $N,\ N-1$ and up to $L=24$ for block $N-2$. Particularly, a clean crossover of $r$ from GOE to Poisson is revealed in block $N-2$ upon raising $L$. Therefore, in accord with the exact scaling trend perceived from the small chains (Figs.~\ref{pic8} and \ref{pic_ni}), on the theoretical side, the scaling analysis based on these perturbative long-chain results (Fig.~\ref{pic_vv}) tentatively confirms our key speculations on the stabilization of the cluster MBL regime and, more importantly, the absence of finite-size drift in the higher-energy section of the phase diagram. For completeness, we also extend the Van Vleck method to the bottom of the phase diagram and consistently recover the convergent trend of $r$ toward GOE in the lowest two blocks $n_{\textrm{max}}=1$ and $2$, verifying the applicability of the method to a wider parameter space. Admittedly, it is worth cautioning that, unlike the top and bottom blocks, for those decisive $n_\textrm{max}$ in the middle of the spectrum, both the block size and the complexity of (\ref{Hdbhprime}) get enhanced enormously such that the Van Vleck algorithm again becomes impractical.

{\color{black} Complementarily, as a useful self check, it would be instructive to directly compare the ED results in Figs.~\ref{pic8} and \ref{pic_ni} with the perturbative results in Fig.~\ref{pic_vv} to potentially evaluate the reliability of the proposed computational scheme based on the Van Vleck theory. To this end, for a set of chosen parameters: $L,\ \mu,\ n_{\rm max}$, we first zoom in the corresponding panel with the specified length $L$ in Fig.~\ref{pic_ni} to identify the target normalized energy $\varepsilon_{n_{\rm max}}$ whose associated eigenstates possess the averaged maximal onsite occupation number within the required range $[n_{\rm max},n_{\rm max}-1)$ but simultaneously being as close to $n_{\rm max}$ as possible when the disorder strength is $\mu$. Subsequently, upon the determination of the target $\varepsilon_{n_{\rm max}},\ \mu,\ L$, we can switch to zoom in the corresponding panel with the same specified length $L$ in Fig.~\ref{pic8} to uniquely pinpoint the value of the target level-spacing ratio from ED, denoted as $r_{\rm ED}$, by its two coordinates $\varepsilon_{n_{\rm max}}$ and $\mu$. Upon specifying $\mu=2J$, this $r_{\rm ED}$ can then be directly contrasted with the value of $r_{\rm pert.}$ read out from Fig.~\ref{pic_vv}. We successfully implement the above procedure and make such a direct comparison in Fig.~\ref{pic_pert_ed}, focusing on the four representative values of $n_{\rm max}=N,\ N-1,\ N-2,\ 1$. From there, it can be observed that ranging from the size $L=8$~up~to~$14$, for each $n_{\rm max}$ studied, the value of $r_{\rm ED}$ and the value of $r_{\rm pert.}$ are quantitatively close to each other. Further, their variation trends under the increase of $L$ are also parallel to each other, suggesting the likely convergence of these two estimates still when extrapolating to the larger sizes.  

Finally, to reinforce the role of the quantum statistics, it is worth emphasizing the difference between the limit of large $n_{\rm max}$, the maximal onsite occupation number, and the limit of large $U$, the interaction strength. First of all, our chosen interaction strength $U=3J$ is not large. Thus, the reason allowing the argument about the blocks to work is not due to the value of $U$, but rather due to the maximal onsite boson occupation number $n_\textrm{max}$. Notice that in our calculations, $n_\textrm{max}=1,\ldots,L/2$. For later notational convenience, let us denote $n_\textrm{max}=\alpha_\textrm{max}L$ where $\alpha_\textrm{max}=1/L,2/L,\ldots,1/2$. Then, for a higher-energy block labelled by say $\alpha_\textrm{max}\approx0.25$, it is easy to estimate that the gap between the $\alpha_\textrm{max}L$ block and the $\alpha_\textrm{max}L+1$ block is about (below always assuming $\mu$ is small)
\beq
\frac{U}{2}(\alpha_\textrm{max}L+1)(\alpha_\textrm{max}L)-\frac{U}{2}\alpha_\textrm{max}L(\alpha_\textrm{max}L-1)=U\alpha_\textrm{max}L.
\label{ualphal}
\eeq
Clearly, this means that it is not $U$ but the nontrivial dependence of $L$ that renders the energy gap between the two adjacent blocks sufficiently large. So, at least, half of the values of $n_\textrm{max}$ from $L/4$ to $L/2$ form the well-separated blocks. Most of these blocks will be within the higher-energy section. Put differently, allowing the multiple occupancy in a given orbital or site is the key of the above analysis. 

By contrast, in the case of spinful fermions such as the spinful Fermi-Hubbard model, we need to concern clusters of doubly-occupied sites. The relevant quantity in this case is the number of such double-occupancies in the chain, denoted as $\tilde{n}$. Then if we use this number $\tilde{n}$ to arrange and label the block, it is easy to find that the gap between blocks $\tilde{n}$ and $\tilde{n}+1$ is about
\beq
U (\tilde{n}+1)-U \tilde{n}=U.
\eeq
Therefore, in the spinful Fermi-Hubbard chain, one needs to require the large value of the onsite repulsion $U$, at the order of $10J$ or more, to observe the potential non-chaotic signature. Physically, this higher-energy section in the spinful-fermion model might correspond to the lower-energy section of the dBH model where the $n_{\rm max}=2$ block locates. Likewise, in the case of spinless fermions, we need to concern clusters of successively-occupied sites. The relevant quantity in this case is the given number of nearest neighbors that have occupancies or the size of such fully-occupied clusters in the chain, denoted as $\tilde{l}$. Then if we use this number $\tilde{l}$ to arrange and label the block, it is easy to find that the gap between blocks $\tilde{l}$ and $\tilde{l}+1$ is about
\beq
V (\tilde{l}+1)-V \tilde{l}=V.
\eeq  
Therefore, in the spinless Fermi-Hubbard chain, one needs to require the large value of the nearest-neighbor repulsion $V$ to observe the potential non-chaotic signature. Physically, this higher-energy section in the spinless-fermion model might also be similar to the lower-energy section of the dBH model.

In summary, from the above analyses, we can see that the key character of the Hilbert-space structure in the higher-energy section of the dBH model is that its local occupation number can be very large, far beyond the single and double occupancies. Although the quantum statistical properties of indistinguishable particles should be derived and examined by exchanging their positions, we know that the fact that there is no limitation on the local occupation number is the physical consequence that the involved particles obey the Bose statistics. Therefore, combine these observations, we are tempted to suggest that the absence of the drift is due to the intrinsic Bose nature of the particles. Further, there is less controversy on the point that the lower-energy section of the dBH chain is qualitatively governed by the spin/hard-core boson/Fermi statistics (see the Sec.~\ref{sec_fermi} below). Then, the observed sharp difference between the dynamical phase diagrams in the lower- and the higher-energy sections, as demonstrated by Figs.~\ref{pic8} and \ref{pic_ni}, naturally hints that the origin of this discrepancy might be traced back to the Fermi-Bose distinction in the emergent and intrinsic particle statistics.

As an aside, the Hilbert-space fragmentation \cite{SalaPollmann} requires that the number of the fragmented blocks should be an exponential function of the system length $L$. However, from Eq.~(\ref{nmax}), we know that the block we identify for the dBH chain is labelled by $n_\textrm{max}=1,2,\ldots,L/2$, which is only a linear function of $L$. Thus, we are hesitated to think that the absence of the drift is the consequence of the Hilbert-space fragmentation. See also the pertinent discussions in Sec.~\ref{secdiscon}.}

\section{The lower-energy limit: mapping to the disordered spin chain model via the algebraic projection method}
\label{sec_fermi}

Contrastingly, due to the energy penalty from the onsite Hubbard repulsion, the lower-energy eigenstates of (\ref{hamdbh}) comprise the state configurations featuring low local boson concentrations. 

To uncover the emergent spin/hard-core boson/fermion degrees of freedom in this bosonic model \cite{Cazalilla,basak2021strongly,huber2011topological}, we illustrate here that the lower-energy portion of the dBH model's dynamical phase diagram (Fig.~\ref{pic8}) can be qualitatively understood by connecting it to the disordered spin chain using the algebraic projection method. To this end, we introduce the projector $\mathcal{P}$ that projects onto the subspace $\mathscr{C}$ subtended by an ensemble of eigenvectors satisfying the hard-core constraint meaning that the onsite boson occupation number on any lattice site is no more than $1$. Likewise, the complement projector $\mathcal{Q}$ that projects onto the supplementary Hilbert space is defined by $\mathcal{Q}=1-\mathcal{P}$. Strictly speaking, within $\mathscr{C}$, the Hubbard $U$ term has no contribution. Therefore, unlike the study of equilibrium ground-state physics, in the algebraic projection approach, the perturbative treatment of the virtual hopping processes [see (\ref{heffp2}) below] is due to the imposition of projectors $\mathcal{P}$ and $\mathcal{Q}$ irrespective of the values of $U$ and $J$. Symbolically, the dBH Hamiltonian can now be written as a matrix,
\beq
\left[{\begin{array}{cc}
  \mathcal{P}H_{\textrm{dBH}}\mathcal{P} & \mathcal{P}H_{\textrm{dBH}}\mathcal{Q}\\
  \mathcal{Q}H_{\textrm{dBH}}\mathcal{P} & \mathcal{Q}H_{\textrm{dBH}}\mathcal{Q}\\
\end{array}}\right].
\eeq
Our intent is to derive an effective Hamiltonian $H_{\textrm{eff}}$ capturing perturbatively the influence of $\mathcal{Q}H_{\textrm{dBH}}\mathcal{Q}$ onto $\mathcal{P}H_{\textrm{dBH}}\mathcal{P}$ via the two off-diagonal couplings $\mathcal{P}H_{\textrm{dBH}}\mathcal{Q}$ and $\mathcal{Q}H_{\textrm{dBH}}\mathcal{P}$. Operationally, such physical processes are encapsulated in the procedure of an approximate block diagonalization of the above matrix. Recall that for a usual $2\times2$ matrix $\Big[{\begin{array}{cc}
    a & b\\
    c & d\\
\end{array}}\Big]$, its two eigenvalues are $\frac{1}{2}(a+d\mp\sqrt{(a-d)^2+4bc})$. Then, under the assumption that the operator norm $||\mathcal{P}H_{\textrm{dBH}}\mathcal{P}||\ll||\mathcal{Q}H_{\textrm{dBH}}\mathcal{Q}||$, it is easy to derive that
\beq
H_{\textrm{eff}}\approx\mathcal{P}H_{\textrm{dBH}}\mathcal{P}-\mathcal{P}H_{\textrm{dBH}}\mathcal{Q}\frac{1}{\mathcal{Q}H_{\textrm{dBH}}\mathcal{Q}}\mathcal{Q}H_{\textrm{dBH}}\mathcal{P},
\label{heff}
\eeq
where we omit the higher-order corrections \cite{lacroix2011introduction}. Alternatively, using resolvent \cite{CohenTannoudji}, one can show that
\beq
[\mathcal{P}H_{\textrm{dBH}}\mathcal{P}-\mathcal{P}H_{\textrm{dBH}}\mathcal{Q}\frac{1}{\mathcal{Q}H_{\textrm{dBH}}\mathcal{Q}}\mathcal{Q}H_{\textrm{dBH}}\mathcal{P}][\mathcal{P}\frac{1}{H_\textrm{dBH}}\mathcal{P}]=1.
\eeq
This instead is an exact identity.

Let us begin with the first term of $H_{\textrm{eff}}$. As usual, the effect of $\mathcal{P}$ can be realized by imposing the hard-core boson constraint, which, in terms of spin-$\frac{1}{2}$ Pauli matrices, can be expressed as the following mapping,
\beq
a_i\rightarrow\sigma^+_i,\ \ \ \ a^\dagger_i\rightarrow\sigma^-_i,\ \ \ \ 1-2n_i\rightarrow\sigma^z_i.
\label{spinmap}
\eeq
{\color{black} Recall the definitions that $\sigma^\pm_i=(\sigma^x_i\pm\textrm{i}\sigma^y_i)/2$. It is important to emphasize that the fermionic statistics would only manifest once the Jordan-Wigner transformation is implemented upon (\ref{spinmap}), as \cite{Cazalilla} did, i.e.,
\begin{align}
\sigma^x_i&=K_i(c^\dagger_i+c_i), \\
\sigma^y_i&=K_i \textrm{i}(c^\dagger_i-c_i), \\
\sigma^z_i&=1-2c^\dagger_ic_i,
\end{align}
where $c^{(\dagger)}_i$ is the fermionic annihilation (creation) operator on site $i$ and the string operator $K_i=\prod^{i-1}_{i'=1}e^{\textrm{i}\pi c^\dagger_{i'}c_{i'}}$.} Thus, under the hard-core boson limit,
\beq
\mathcal{P}H_{\textrm{dBH}}\mathcal{P}=-\frac{J}{2}\sum^{L}_{i=1}(\sigma^x_i\sigma^x_{i+1}+\sigma^y_i\sigma^y_{i+1})+\sum^L_{i=1}\frac{\mu_i}{2}(1-\sigma^z_i).
\label{heffp1}
\eeq
It is ready to recognize that $\mathcal{P}H_{\textrm{dBH}}\mathcal{P}$ is a disordered $X\!X$ spin chain, which is Anderson localized in 1D. Meanwhile, upon increasing $U$ to infinity, $\mathcal{P}H_{\textrm{dBH}}\mathcal{P}$ would be the only remaining term in $H_{\textrm{dBH}}$ and the dBH model in this limit becomes a free-fermion chain. In this sense, the genuine many-body interaction effects in $H_{\textrm{eff}}$ shall arise for moderate $U$ and they stem mainly from the second as well as those omitted higher-order correction terms in (\ref{heff}).

Next, for the second term of $H_{\textrm{eff}}$, we would only consider the second-order virtual hopping processes, yielding
\begin{widetext}
\begin{align}
-\mathcal{P}H_{\textrm{dBH}}\mathcal{Q}\frac{1}{\mathcal{Q}H_{\textrm{dBH}}\mathcal{Q}}\mathcal{Q}H_{\textrm{dBH}}\mathcal{P}&\approx-J^2\sum^L_{i=1}\mathcal{P}(a^\dagger_{i-1}+a^\dagger_{i+1})a_i\frac{n_i-1}{U+\sum\limits^L_{j=1}\mu_jn_j}a^\dagger_i(a_{i-1}+a_{i+1})\mathcal{P} \nonumber \\
&\approx\sum^L_{i=1}(\sigma^-_{i-1}+\sigma^-_{i+1})\sigma^+_{i+1}\frac{(-4J^2)}{2U+4\mu_i+\sum\limits_{j\neq i,i+1}\mu_j(1-\sigma^z_j)} \nonumber \\
&+\sum^L_{i=1}(\sigma^-_{i-1}+\sigma^-_{i+1})\sigma^+_{i-1}\frac{(-4J^2)}{2U+4\mu_i+\sum\limits_{j\neq i,i-1}\mu_j(1-\sigma^z_j)},
\label{heffp2}
\end{align}
\end{widetext}
where in middle steps we replace $\mathcal{Q}$ by the equivalent operator $n_i-1$ and then $\mathcal{P}$ is removed after invoking (\ref{spinmap}).

Collectively, up to the second-order perturbation, the effective spin Hamiltonian $H_{\textrm{eff}}$ capable of qualitatively describing the lower-energy portion of the dynamical phase diagram of the dBH model is given by summing up expressions (\ref{heffp1}) and (\ref{heffp2}). For the case of $\mu_i=0$, this result was obtained before by \cite{Cazalilla}. However, it is interesting to notice that when $\mu_i\neq0$ and under the lower-energy limit, besides the resulting diagonal disordered $\sigma^z$-fields in the noninteracting part of $H_{\textrm{eff}}$, i.e., (\ref{heffp1}), the leading off-diagonal many-body interacting part of $H_{\textrm{eff}}$, i.e., (\ref{heffp2}), becomes randomized as well, whose disorder strength is further characterized by a dynamical dependence on the particular $\sigma^z$-configuration of the acted basis state.

As model (\ref{heffp1}) is Anderson localized, the significant finite-size drift seen in Figs.~\ref{pic8} and \ref{pic_ni} at $U=3J$ that obscures the identification of MBL in the dBH chain shall originate from the revealed longer-range multi-spin interactions \cite{Molignini_arxiv} in Eq.~(\ref{heffp2}). In light of the importance of the dBH model in the experiments \cite{LukinGreiner,leonard2023probing}, the derived disordered interacting spin model (\ref{heffp2}) may itself be an intriguing model for the future study of MBL in the bosonic systems.

In essence, the above analysis indicates that the lower-energy nonequilibrium physics of the dBH model is largely governed by the particles' emergent spin or Fermi statistics.

\section{Discussion and conclusion} \label{secdiscon}

{\color{black} For completeness, here we briefly comment on the main difference between the present work and some representative works of the earlier literature on the dBH chain. In the disorder-free case, Ref.~\cite{Carleo} targeted the lower-energy section of the dynamical phase diagram of the clean BH model, and its main findings could be understood by the mapping to a spin chain model using the algebraic projection method. The nonergodic behavior observed in Ref.~\cite{Carleo} had nothing to do with the Van Vleck general formula (\ref{Hdbhprime}) derived in the higher-energy section. The nonergodicity in Ref.~\cite{Carleo} requires large $U=10J$, so it hinted at some signature of the Hilbert-space fragmentation \cite{SalaPollmann} in the lower-energy section of the dynamical phase diagram due to the strong Hubbard repulsion \cite{Honda,Kollath}. In sharp contrast, we show in Fig.~8 of Ref.~\cite{jiechen2023} that for $U=3J,\ N=\frac{L}{2}$, the clean BH chain in the higher-energy section (corresponding to the upper part of the vertical line at $\mu=0$ in Fig.~\ref{pic8}) exhibits the dynamical behaviors consistent with the phenomenon of prethermalization or thermalization. This observation contrasts with the nonergodicity seen in Ref.~\cite{Carleo}, thus precludes the role of the Hilbert-space fragmentation in the parameter space we are considering in this work, and further cements the distinction between our work and Ref.~\cite{Carleo}. Moreover, this prethermal or thermal regime at $\mu=0$ also contrasts with the cluster MBL regime realized at small $\mu$ in Fig.~\ref{pic8}. In the disordered case, our work is partly inspired by Ref.~\cite{Sierant_2018} where the energy-resolved dynamical phase diagram of the dBH chain was mapped out but at a different filling fraction. However, the focus of Ref.~\cite{Sierant_2018}, like many others \cite{yao2020many,orell2019probing,hopjan2020many,Sarkar}, was zoomed in the lower-energy section as well. The higher-energy section of the dynamical phase diagram of the dBH chain is thus comparatively less explored so far. In addition, the present work only addresses the situation where the disorder is added upon the onsite chemical potential. Ref.~\cite{Sierant2017} discussed the interesting alternative that the disorder could be on the strength of the many-body interaction term. It would be intriguing to examine the potential MBL physics of bosons from the randomized Hubbard repulsion along the similar lines of Ref.~\cite{Sierant2017Poland}. Moreover, the generalization of our results to the non-Hermitian disordered systems \cite{Molignini_skin} might be another interesting direction to pursue.}

To summarize, like dimensionality and symmetry, particle statistics can influence the eigenstate matter formation. Through the introduction of the Van Vleck perturbation theory tailored for handling the clustering structures due to Bose statistics, we are tempted to speculate on the absence of the finite-size drift and the robustness of the cluster MBL regime in the higher-energy section of the dBH model's dynamical phase diagram. On the contrary, via the algebraic projection approach, the persisting finite-size drift and the successive enlargement of the thermal phase in the lower-energy section of the same phase diagram are then partially explained by invoking the emergent spin/Fermi statistics so as to map this bosonic model onto a disordered spin chain model. These disparate scaling behaviors hint that the Bose-Fermi distinction in the particle statistics may delineate a mobility edge in between the cluster MBL regime in the higher-energy section and the thermal phase in the lower-energy section for the dBH type chains.

It is noteworthy that the scaling of the entanglement entropy and its quantum quench dynamics have been studied for the dBH model in the two accompanying papers \cite{chen2024eigenstate,jiechen2023}. The results drawn from there are consistent with the spectral results presented in this work.

\section{Acknowledgements}

We thank the anonymous Referees for their insightful comments and constructive suggestions. We thank Z. Cai for the useful discussions. J.~C. and X.~W. were supported by MOST2022YFA1402701 and the NSFC Grant No.~11974244. C.~C. was supported by a start-up fund from SJTU and the sponsorship from Yangyang Development fund.

\bibliography{dbh}

\begin{thebibliography}{46}%
\makeatletter
\providecommand \@ifxundefined [1]{%
 \@ifx{#1\undefined}
}%
\providecommand \@ifnum [1]{%
 \ifnum #1\expandafter \@firstoftwo
 \else \expandafter \@secondoftwo
 \fi
}%
\providecommand \@ifx [1]{%
 \ifx #1\expandafter \@firstoftwo
 \else \expandafter \@secondoftwo
 \fi
}%
\providecommand \natexlab [1]{#1}%
\providecommand \enquote  [1]{``#1''}%
\providecommand \bibnamefont  [1]{#1}%
\providecommand \bibfnamefont [1]{#1}%
\providecommand \citenamefont [1]{#1}%
\providecommand \href@noop [0]{\@secondoftwo}%
\providecommand \href [0]{\begingroup \@sanitize@url \@href}%
\providecommand \@href[1]{\@@startlink{#1}\@@href}%
\providecommand \@@href[1]{\endgroup#1\@@endlink}%
\providecommand \@sanitize@url [0]{\catcode `\\12\catcode `\$12\catcode
  `\&12\catcode `\#12\catcode `\^12\catcode `\_12\catcode `\%12\relax}%
\providecommand \@@startlink[1]{}%
\providecommand \@@endlink[0]{}%
\providecommand \url  [0]{\begingroup\@sanitize@url \@url }%
\providecommand \@url [1]{\endgroup\@href {#1}{\urlprefix }}%
\providecommand \urlprefix  [0]{URL }%
\providecommand \Eprint [0]{\href }%
\providecommand \doibase [0]{https://doi.org/}%
\providecommand \selectlanguage [0]{\@gobble}%
\providecommand \bibinfo  [0]{\@secondoftwo}%
\providecommand \bibfield  [0]{\@secondoftwo}%
\providecommand \translation [1]{[#1]}%
\providecommand \BibitemOpen [0]{}%
\providecommand \bibitemStop [0]{}%
\providecommand \bibitemNoStop [0]{.\EOS\space}%
\providecommand \EOS [0]{\spacefactor3000\relax}%
\providecommand \BibitemShut  [1]{\csname bibitem#1\endcsname}%
\let\auto@bib@innerbib\@empty
\bibitem [{\citenamefont {Morningstar}\ \emph {et~al.}(2022)\citenamefont
  {Morningstar}, \citenamefont {Colmenarez}, \citenamefont {Khemani},
  \citenamefont {Luitz},\ and\ \citenamefont
  {Huse}}]{morningstar2022avalanches}%
  \BibitemOpen
  \bibfield  {author} {\bibinfo {author} {\bibfnamefont {A.}~\bibnamefont
  {Morningstar}}, \bibinfo {author} {\bibfnamefont {L.}~\bibnamefont
  {Colmenarez}}, \bibinfo {author} {\bibfnamefont {V.}~\bibnamefont {Khemani}},
  \bibinfo {author} {\bibfnamefont {D.~J.}\ \bibnamefont {Luitz}},\ and\
  \bibinfo {author} {\bibfnamefont {D.~A.}\ \bibnamefont {Huse}},\ }\bibfield
  {title} {\bibinfo {title} {{Avalanches and many-body resonances in many-body
  localized systems}},\ }\href
  {https://journals.aps.org/prb/abstract/10.1103/PhysRevB.105.174205}
  {\bibfield  {journal} {\bibinfo  {journal} {Phys. Rev. B}\ }\textbf {\bibinfo
  {volume} {105}},\ \bibinfo {pages} {174205} (\bibinfo {year}
  {2022})}\BibitemShut {NoStop}%
\bibitem [{\citenamefont {\ifmmode~\check{S}\else \v{S}\fi{}untajs}\ \emph
  {et~al.}(2020)\citenamefont {\ifmmode~\check{S}\else \v{S}\fi{}untajs},
  \citenamefont {Bon\ifmmode~\check{c}\else \v{c}\fi{}a}, \citenamefont
  {Prosen},\ and\ \citenamefont {Vidmar}}]{Suntajs}%
  \BibitemOpen
  \bibfield  {author} {\bibinfo {author} {\bibfnamefont {J.}~\bibnamefont
  {\ifmmode~\check{S}\else \v{S}\fi{}untajs}}, \bibinfo {author} {\bibfnamefont
  {J.}~\bibnamefont {Bon\ifmmode~\check{c}\else \v{c}\fi{}a}}, \bibinfo
  {author} {\bibfnamefont {T.}~\bibnamefont {Prosen}},\ and\ \bibinfo {author}
  {\bibfnamefont {L.}~\bibnamefont {Vidmar}},\ }\bibfield  {title} {\bibinfo
  {title} {{Quantum chaos challenges many-body localization}},\ }\href
  {https://doi.org/10.1103/PhysRevE.102.062144} {\bibfield  {journal} {\bibinfo
   {journal} {Phys. Rev. E}\ }\textbf {\bibinfo {volume} {102}},\ \bibinfo
  {pages} {062144} (\bibinfo {year} {2020})}\BibitemShut {NoStop}%
\bibitem [{\citenamefont {Sierant}\ \emph {et~al.}(2025)\citenamefont
  {Sierant}, \citenamefont {Lewenstein}, \citenamefont {Scardicchio},
  \citenamefont {Vidmar},\ and\ \citenamefont {Zakrzewski}}]{Sierant2024}%
  \BibitemOpen
  \bibfield  {author} {\bibinfo {author} {\bibfnamefont {P.}~\bibnamefont
  {Sierant}}, \bibinfo {author} {\bibfnamefont {M.}~\bibnamefont {Lewenstein}},
  \bibinfo {author} {\bibfnamefont {A.}~\bibnamefont {Scardicchio}}, \bibinfo
  {author} {\bibfnamefont {L.}~\bibnamefont {Vidmar}},\ and\ \bibinfo {author}
  {\bibfnamefont {J.}~\bibnamefont {Zakrzewski}},\ }\bibfield  {title}
  {\bibinfo {title} {{Many-body localization in the age of classical
  computing}},\ }\href {https://doi.org/10.1088/1361-6633/ad9756} {\bibfield
  {journal} {\bibinfo  {journal} {Rep. Prog. Phys.}\ }\textbf {\bibinfo
  {volume} {88}},\ \bibinfo {pages} {026502} (\bibinfo {year}
  {2025})}\BibitemShut {NoStop}%
\bibitem [{\citenamefont {Kiefer-Emmanouilidis}\ \emph
  {et~al.}(2020)\citenamefont {Kiefer-Emmanouilidis}, \citenamefont {Unanyan},
  \citenamefont {Fleischhauer},\ and\ \citenamefont
  {Sirker}}]{KieferEmmanouilidisPRL}%
  \BibitemOpen
  \bibfield  {author} {\bibinfo {author} {\bibfnamefont {M.}~\bibnamefont
  {Kiefer-Emmanouilidis}}, \bibinfo {author} {\bibfnamefont {R.}~\bibnamefont
  {Unanyan}}, \bibinfo {author} {\bibfnamefont {M.}~\bibnamefont
  {Fleischhauer}},\ and\ \bibinfo {author} {\bibfnamefont {J.}~\bibnamefont
  {Sirker}},\ }\bibfield  {title} {\bibinfo {title} {{Evidence for Unbounded
  Growth of the Number Entropy in Many-Body Localized Phases}},\ }\href
  {https://doi.org/10.1103/PhysRevLett.124.243601} {\bibfield  {journal}
  {\bibinfo  {journal} {Phys. Rev. Lett.}\ }\textbf {\bibinfo {volume} {124}},\
  \bibinfo {pages} {243601} (\bibinfo {year} {2020})}\BibitemShut {NoStop}%
\bibitem [{\citenamefont {{\v{S}}untajs}\ \emph {et~al.}(2020)\citenamefont
  {{\v{S}}untajs}, \citenamefont {Bon{\v{c}}a}, \citenamefont {Prosen},\ and\
  \citenamefont {Vidmar}}]{vsuntajs2020ergodicity}%
  \BibitemOpen
  \bibfield  {author} {\bibinfo {author} {\bibfnamefont {J.}~\bibnamefont
  {{\v{S}}untajs}}, \bibinfo {author} {\bibfnamefont {J.}~\bibnamefont
  {Bon{\v{c}}a}}, \bibinfo {author} {\bibfnamefont {T.}~\bibnamefont
  {Prosen}},\ and\ \bibinfo {author} {\bibfnamefont {L.}~\bibnamefont
  {Vidmar}},\ }\bibfield  {title} {\bibinfo {title} {{Ergodicity breaking
  transition in finite disordered spin chains}},\ }\href
  {https://journals.aps.org/prb/abstract/10.1103/PhysRevB.102.064207}
  {\bibfield  {journal} {\bibinfo  {journal} {Phys. Rev. B}\ }\textbf {\bibinfo
  {volume} {102}},\ \bibinfo {pages} {064207} (\bibinfo {year}
  {2020})}\BibitemShut {NoStop}%
\bibitem [{\citenamefont {Long}\ \emph {et~al.}(2023)\citenamefont {Long},
  \citenamefont {Crowley}, \citenamefont {Khemani},\ and\ \citenamefont
  {Chandran}}]{long2023phenomenology}%
  \BibitemOpen
  \bibfield  {author} {\bibinfo {author} {\bibfnamefont {D.~M.}\ \bibnamefont
  {Long}}, \bibinfo {author} {\bibfnamefont {P.~J.}\ \bibnamefont {Crowley}},
  \bibinfo {author} {\bibfnamefont {V.}~\bibnamefont {Khemani}},\ and\ \bibinfo
  {author} {\bibfnamefont {A.}~\bibnamefont {Chandran}},\ }\bibfield  {title}
  {\bibinfo {title} {{Phenomenology of the Prethermal Many-Body Localized
  Regime}},\ }\href
  {https://journals.aps.org/prl/abstract/10.1103/PhysRevLett.131.106301}
  {\bibfield  {journal} {\bibinfo  {journal} {Phys. Rev. Lett.}\ }\textbf
  {\bibinfo {volume} {131}},\ \bibinfo {pages} {106301} (\bibinfo {year}
  {2023})}\BibitemShut {NoStop}%
\bibitem [{\citenamefont {Aceituno~Ch{\'a}vez}\ \emph
  {et~al.}(2024)\citenamefont {Aceituno~Ch{\'a}vez}, \citenamefont {Artiaco},
  \citenamefont {Klein~Kvorning}, \citenamefont {Herviou},\ and\ \citenamefont
  {Bardarson}}]{aceituno2024ultraslow}%
  \BibitemOpen
  \bibfield  {author} {\bibinfo {author} {\bibfnamefont {D.}~\bibnamefont
  {Aceituno~Ch{\'a}vez}}, \bibinfo {author} {\bibfnamefont {C.}~\bibnamefont
  {Artiaco}}, \bibinfo {author} {\bibfnamefont {T.}~\bibnamefont
  {Klein~Kvorning}}, \bibinfo {author} {\bibfnamefont {L.}~\bibnamefont
  {Herviou}},\ and\ \bibinfo {author} {\bibfnamefont {J.~H.}\ \bibnamefont
  {Bardarson}},\ }\bibfield  {title} {\bibinfo {title} {{Ultraslow Growth of
  Number Entropy in an $\ell$-Bit Model of Many-Body Localization}},\ }\href
  {https://journals.aps.org/prl/abstract/10.1103/PhysRevLett.133.126502}
  {\bibfield  {journal} {\bibinfo  {journal} {Phys. Rev. Lett.}\ }\textbf
  {\bibinfo {volume} {133}},\ \bibinfo {pages} {126502} (\bibinfo {year}
  {2024})}\BibitemShut {NoStop}%
\bibitem [{\citenamefont {Luitz}\ \emph {et~al.}(2015)\citenamefont {Luitz},
  \citenamefont {Laflorencie},\ and\ \citenamefont {Alet}}]{Luitz}%
  \BibitemOpen
  \bibfield  {author} {\bibinfo {author} {\bibfnamefont {D.~J.}\ \bibnamefont
  {Luitz}}, \bibinfo {author} {\bibfnamefont {N.}~\bibnamefont {Laflorencie}},\
  and\ \bibinfo {author} {\bibfnamefont {F.}~\bibnamefont {Alet}},\ }\bibfield
  {title} {\bibinfo {title} {{Many-body localization edge in the random-field
  Heisenberg chain}},\ }\href {https://doi.org/10.1103/PhysRevB.91.081103}
  {\bibfield  {journal} {\bibinfo  {journal} {Phys. Rev. B}\ }\textbf {\bibinfo
  {volume} {91}},\ \bibinfo {pages} {081103} (\bibinfo {year}
  {2015})}\BibitemShut {NoStop}%
\bibitem [{\citenamefont {Sierant}\ \emph {et~al.}(2020)\citenamefont
  {Sierant}, \citenamefont {Lewenstein},\ and\ \citenamefont
  {Zakrzewski}}]{sierant2020polynomially}%
  \BibitemOpen
  \bibfield  {author} {\bibinfo {author} {\bibfnamefont {P.}~\bibnamefont
  {Sierant}}, \bibinfo {author} {\bibfnamefont {M.}~\bibnamefont
  {Lewenstein}},\ and\ \bibinfo {author} {\bibfnamefont {J.}~\bibnamefont
  {Zakrzewski}},\ }\bibfield  {title} {\bibinfo {title} {{Polynomially Filtered
  Exact Diagonalization Approach to Many-Body Localization}},\ }\href
  {https://journals.aps.org/prl/abstract/10.1103/PhysRevLett.125.156601}
  {\bibfield  {journal} {\bibinfo  {journal} {Phys. Rev. Lett.}\ }\textbf
  {\bibinfo {volume} {125}},\ \bibinfo {pages} {156601} (\bibinfo {year}
  {2020})}\BibitemShut {NoStop}%
\bibitem [{\citenamefont {Colbois}\ \emph {et~al.}(2024)\citenamefont
  {Colbois}, \citenamefont {Alet},\ and\ \citenamefont
  {Laflorencie}}]{colbois2024interaction}%
  \BibitemOpen
  \bibfield  {author} {\bibinfo {author} {\bibfnamefont {J.}~\bibnamefont
  {Colbois}}, \bibinfo {author} {\bibfnamefont {F.}~\bibnamefont {Alet}},\ and\
  \bibinfo {author} {\bibfnamefont {N.}~\bibnamefont {Laflorencie}},\
  }\bibfield  {title} {\bibinfo {title} {{Interaction-Driven Instabilities in
  the Random-Field $X\!X\!Z$ Chain}},\ }\href
  {https://journals.aps.org/prl/abstract/10.1103/PhysRevLett.133.116502}
  {\bibfield  {journal} {\bibinfo  {journal} {Phys. Rev. Lett.}\ }\textbf
  {\bibinfo {volume} {133}},\ \bibinfo {pages} {116502} (\bibinfo {year}
  {2024})}\BibitemShut {NoStop}%
\bibitem [{\citenamefont {Hopjan}\ and\ \citenamefont
  {Heidrich-Meisner}(2020)}]{hopjan2020many}%
  \BibitemOpen
  \bibfield  {author} {\bibinfo {author} {\bibfnamefont {M.}~\bibnamefont
  {Hopjan}}\ and\ \bibinfo {author} {\bibfnamefont {F.}~\bibnamefont
  {Heidrich-Meisner}},\ }\bibfield  {title} {\bibinfo {title} {{Many-body
  localization from a one-particle perspective in the disordered
  one-dimensional Bose-Hubbard model}},\ }\href
  {https://journals.aps.org/pra/abstract/10.1103/PhysRevA.101.063617}
  {\bibfield  {journal} {\bibinfo  {journal} {Phys. Rev. A}\ }\textbf {\bibinfo
  {volume} {101}},\ \bibinfo {pages} {063617} (\bibinfo {year}
  {2020})}\BibitemShut {NoStop}%
\bibitem [{\citenamefont {Orell}\ \emph {et~al.}(2019)\citenamefont {Orell},
  \citenamefont {Michailidis}, \citenamefont {Serbyn},\ and\ \citenamefont
  {Silveri}}]{orell2019probing}%
  \BibitemOpen
  \bibfield  {author} {\bibinfo {author} {\bibfnamefont {T.}~\bibnamefont
  {Orell}}, \bibinfo {author} {\bibfnamefont {A.~A.}\ \bibnamefont
  {Michailidis}}, \bibinfo {author} {\bibfnamefont {M.}~\bibnamefont
  {Serbyn}},\ and\ \bibinfo {author} {\bibfnamefont {M.}~\bibnamefont
  {Silveri}},\ }\bibfield  {title} {\bibinfo {title} {{Probing the many-body
  localization phase transition with superconducting circuits}},\ }\href
  {https://journals.aps.org/prb/abstract/10.1103/PhysRevB.100.134504}
  {\bibfield  {journal} {\bibinfo  {journal} {Phys. Rev. B}\ }\textbf {\bibinfo
  {volume} {100}},\ \bibinfo {pages} {134504} (\bibinfo {year}
  {2019})}\BibitemShut {NoStop}%
\bibitem [{\citenamefont {Sierant}\ and\ \citenamefont
  {Zakrzewski}(2018)}]{Sierant_2018}%
  \BibitemOpen
  \bibfield  {author} {\bibinfo {author} {\bibfnamefont {P.}~\bibnamefont
  {Sierant}}\ and\ \bibinfo {author} {\bibfnamefont {J.}~\bibnamefont
  {Zakrzewski}},\ }\bibfield  {title} {\bibinfo {title} {{Many-body
  localization of bosons in optical lattices}},\ }\href
  {https://doi.org/10.1088/1367-2630/aabb17} {\bibfield  {journal} {\bibinfo
  {journal} {New J. Phys.}\ }\textbf {\bibinfo {volume} {20}},\ \bibinfo
  {pages} {043032} (\bibinfo {year} {2018})}\BibitemShut {NoStop}%
\bibitem [{\citenamefont {Yao}\ and\ \citenamefont
  {Zakrzewski}(2020)}]{yao2020many}%
  \BibitemOpen
  \bibfield  {author} {\bibinfo {author} {\bibfnamefont {R.}~\bibnamefont
  {Yao}}\ and\ \bibinfo {author} {\bibfnamefont {J.}~\bibnamefont
  {Zakrzewski}},\ }\bibfield  {title} {\bibinfo {title} {{Many-body
  localization in the Bose-Hubbard model: Evidence for mobility edge}},\ }\href
  {https://journals.aps.org/prb/abstract/10.1103/PhysRevB.102.014310}
  {\bibfield  {journal} {\bibinfo  {journal} {Phys. Rev. B}\ }\textbf {\bibinfo
  {volume} {102}},\ \bibinfo {pages} {014310} (\bibinfo {year}
  {2020})}\BibitemShut {NoStop}%
\bibitem [{\citenamefont {Krause}\ \emph {et~al.}(2021)\citenamefont {Krause},
  \citenamefont {Pellegrin}, \citenamefont {Brouwer}, \citenamefont {Abanin},\
  and\ \citenamefont {Filippone}}]{Filippone}%
  \BibitemOpen
  \bibfield  {author} {\bibinfo {author} {\bibfnamefont {U.}~\bibnamefont
  {Krause}}, \bibinfo {author} {\bibfnamefont {T.}~\bibnamefont {Pellegrin}},
  \bibinfo {author} {\bibfnamefont {P.~W.}\ \bibnamefont {Brouwer}}, \bibinfo
  {author} {\bibfnamefont {D.~A.}\ \bibnamefont {Abanin}},\ and\ \bibinfo
  {author} {\bibfnamefont {M.}~\bibnamefont {Filippone}},\ }\bibfield  {title}
  {\bibinfo {title} {{Nucleation of Ergodicity by a Single Mobile Impurity in
  Supercooled Insulators}},\ }\href
  {https://doi.org/10.1103/PhysRevLett.126.030603} {\bibfield  {journal}
  {\bibinfo  {journal} {Phys. Rev. Lett.}\ }\textbf {\bibinfo {volume} {126}},\
  \bibinfo {pages} {030603} (\bibinfo {year} {2021})}\BibitemShut {NoStop}%
\bibitem [{\citenamefont {Molignini}\ and\ \citenamefont
  {Chakrabarti}(2025)}]{Molignini_boson}%
  \BibitemOpen
  \bibfield  {author} {\bibinfo {author} {\bibfnamefont {P.}~\bibnamefont
  {Molignini}}\ and\ \bibinfo {author} {\bibfnamefont {B.}~\bibnamefont
  {Chakrabarti}},\ }\bibfield  {title} {\bibinfo {title} {Stability of dipolar
  bosons in a quasiperiodic potential},\ }\href
  {https://doi.org/10.1103/PhysRevResearch.7.023237} {\bibfield  {journal}
  {\bibinfo  {journal} {Phys. Rev. Res.}\ }\textbf {\bibinfo {volume} {7}},\
  \bibinfo {pages} {023237} (\bibinfo {year} {2025})}\BibitemShut {NoStop}%
\bibitem [{\citenamefont {Lukin}\ \emph {et~al.}(2019)\citenamefont {Lukin},
  \citenamefont {Rispoli}, \citenamefont {Schittko}, \citenamefont {Tai},
  \citenamefont {Kaufman}, \citenamefont {Choi}, \citenamefont {Khemani},
  \citenamefont {L\'eonard},\ and\ \citenamefont {Greiner}}]{LukinGreiner}%
  \BibitemOpen
  \bibfield  {author} {\bibinfo {author} {\bibfnamefont {A.}~\bibnamefont
  {Lukin}}, \bibinfo {author} {\bibfnamefont {M.}~\bibnamefont {Rispoli}},
  \bibinfo {author} {\bibfnamefont {R.}~\bibnamefont {Schittko}}, \bibinfo
  {author} {\bibfnamefont {M.~E.}\ \bibnamefont {Tai}}, \bibinfo {author}
  {\bibfnamefont {A.~M.}\ \bibnamefont {Kaufman}}, \bibinfo {author}
  {\bibfnamefont {S.}~\bibnamefont {Choi}}, \bibinfo {author} {\bibfnamefont
  {V.}~\bibnamefont {Khemani}}, \bibinfo {author} {\bibfnamefont
  {J.}~\bibnamefont {L\'eonard}},\ and\ \bibinfo {author} {\bibfnamefont
  {M.}~\bibnamefont {Greiner}},\ }\bibfield  {title} {\bibinfo {title}
  {{Probing entanglement in a many-body-localized system}},\ }\href
  {https://science.sciencemag.org/content/364/6437/256.full} {\bibfield
  {journal} {\bibinfo  {journal} {Science}\ }\textbf {\bibinfo {volume}
  {364}},\ \bibinfo {pages} {256} (\bibinfo {year} {2019})}\BibitemShut
  {NoStop}%
\bibitem [{\citenamefont {L{\'e}onard}\ \emph {et~al.}(2023)\citenamefont
  {L{\'e}onard}, \citenamefont {Kim}, \citenamefont {Rispoli}, \citenamefont
  {Lukin}, \citenamefont {Schittko}, \citenamefont {Kwan}, \citenamefont
  {Demler}, \citenamefont {Sels},\ and\ \citenamefont
  {Greiner}}]{leonard2023probing}%
  \BibitemOpen
  \bibfield  {author} {\bibinfo {author} {\bibfnamefont {J.}~\bibnamefont
  {L{\'e}onard}}, \bibinfo {author} {\bibfnamefont {S.}~\bibnamefont {Kim}},
  \bibinfo {author} {\bibfnamefont {M.}~\bibnamefont {Rispoli}}, \bibinfo
  {author} {\bibfnamefont {A.}~\bibnamefont {Lukin}}, \bibinfo {author}
  {\bibfnamefont {R.}~\bibnamefont {Schittko}}, \bibinfo {author}
  {\bibfnamefont {J.}~\bibnamefont {Kwan}}, \bibinfo {author} {\bibfnamefont
  {E.}~\bibnamefont {Demler}}, \bibinfo {author} {\bibfnamefont
  {D.}~\bibnamefont {Sels}},\ and\ \bibinfo {author} {\bibfnamefont
  {M.}~\bibnamefont {Greiner}},\ }\bibfield  {title} {\bibinfo {title}
  {{Probing the onset of quantum avalanches in a many-body localized system}},\
  }\href {https://www.nature.com/articles/s41567-022-01887-3} {\bibfield
  {journal} {\bibinfo  {journal} {Nat. Phys.}\ }\textbf {\bibinfo {volume}
  {19}},\ \bibinfo {pages} {481} (\bibinfo {year} {2023})}\BibitemShut
  {NoStop}%
\bibitem [{\citenamefont {Yan}\ \emph {et~al.}(2017{\natexlab{a}})\citenamefont
  {Yan}, \citenamefont {Hui}, \citenamefont {Rigol},\ and\ \citenamefont
  {Scarola}}]{YanPRL}%
  \BibitemOpen
  \bibfield  {author} {\bibinfo {author} {\bibfnamefont {M.}~\bibnamefont
  {Yan}}, \bibinfo {author} {\bibfnamefont {H.-Y.}\ \bibnamefont {Hui}},
  \bibinfo {author} {\bibfnamefont {M.}~\bibnamefont {Rigol}},\ and\ \bibinfo
  {author} {\bibfnamefont {V.~W.}\ \bibnamefont {Scarola}},\ }\bibfield
  {title} {\bibinfo {title} {{Equilibration Dynamics of Strongly Interacting
  Bosons in 2D Lattices with Disorder}},\ }\href
  {https://doi.org/10.1103/PhysRevLett.119.073002} {\bibfield  {journal}
  {\bibinfo  {journal} {Phys. Rev. Lett.}\ }\textbf {\bibinfo {volume} {119}},\
  \bibinfo {pages} {073002} (\bibinfo {year} {2017}{\natexlab{a}})}\BibitemShut
  {NoStop}%
\bibitem [{\citenamefont {Yan}\ \emph {et~al.}(2017{\natexlab{b}})\citenamefont
  {Yan}, \citenamefont {Hui},\ and\ \citenamefont {Scarola}}]{YanPRA}%
  \BibitemOpen
  \bibfield  {author} {\bibinfo {author} {\bibfnamefont {M.}~\bibnamefont
  {Yan}}, \bibinfo {author} {\bibfnamefont {H.-Y.}\ \bibnamefont {Hui}},\ and\
  \bibinfo {author} {\bibfnamefont {V.~W.}\ \bibnamefont {Scarola}},\
  }\bibfield  {title} {\bibinfo {title} {{Dynamics of disordered states in the
  Bose-Hubbard model with confinement}},\ }\href
  {https://doi.org/10.1103/PhysRevA.95.053624} {\bibfield  {journal} {\bibinfo
  {journal} {Phys. Rev. A}\ }\textbf {\bibinfo {volume} {95}},\ \bibinfo
  {pages} {053624} (\bibinfo {year} {2017}{\natexlab{b}})}\BibitemShut
  {NoStop}%
\bibitem [{\citenamefont {Russ}\ \emph {et~al.}(2025)\citenamefont {Russ},
  \citenamefont {Yan}, \citenamefont {Kowalski}, \citenamefont {Wadleigh},
  \citenamefont {Scarola},\ and\ \citenamefont {DeMarco}}]{YanarXiv}%
  \BibitemOpen
  \bibfield  {author} {\bibinfo {author} {\bibfnamefont {P.}~\bibnamefont
  {Russ}}, \bibinfo {author} {\bibfnamefont {M.}~\bibnamefont {Yan}}, \bibinfo
  {author} {\bibfnamefont {N.}~\bibnamefont {Kowalski}}, \bibinfo {author}
  {\bibfnamefont {L.}~\bibnamefont {Wadleigh}}, \bibinfo {author}
  {\bibfnamefont {V.~W.}\ \bibnamefont {Scarola}},\ and\ \bibinfo {author}
  {\bibfnamefont {B.}~\bibnamefont {DeMarco}},\ }\bibfield  {title} {\bibinfo
  {title} {{Compressibility measurement of the thermal MI--BG transition in an
  optical lattice}},\ }\href {https://arxiv.org/abs/2506.16466} {\bibfield
  {journal} {\bibinfo  {journal} {arXiv:2506.16466}\ } (\bibinfo {year}
  {2025})}\BibitemShut {NoStop}%
\bibitem [{\citenamefont {Van~Vleck}(1929)}]{VanVleck}%
  \BibitemOpen
  \bibfield  {author} {\bibinfo {author} {\bibfnamefont {J.~H.}\ \bibnamefont
  {Van~Vleck}},\ }\bibfield  {title} {\bibinfo {title} {{On
  $\ensuremath{\sigma}$-Type Doubling and Electron Spin in the Spectra of
  Diatomic Molecules}},\ }\href {https://doi.org/10.1103/PhysRev.33.467}
  {\bibfield  {journal} {\bibinfo  {journal} {Phys. Rev.}\ }\textbf {\bibinfo
  {volume} {33}},\ \bibinfo {pages} {467} (\bibinfo {year} {1929})}\BibitemShut
  {NoStop}%
\bibitem [{\citenamefont {Cohen-Tannoudji}\ \emph {et~al.}(1998)\citenamefont
  {Cohen-Tannoudji}, \citenamefont {Dupont-Roc},\ and\ \citenamefont
  {Grynberg}}]{CohenTannoudji}%
  \BibitemOpen
  \bibfield  {author} {\bibinfo {author} {\bibfnamefont {C.}~\bibnamefont
  {Cohen-Tannoudji}}, \bibinfo {author} {\bibfnamefont {J.}~\bibnamefont
  {Dupont-Roc}},\ and\ \bibinfo {author} {\bibfnamefont {G.}~\bibnamefont
  {Grynberg}},\ }\href@noop {} {\emph {\bibinfo {title} {{Atom-Photon
  Interactions}}}}\ (\bibinfo  {publisher} {John Wiley \& Sons},\ \bibinfo
  {year} {1998})\BibitemShut {NoStop}%
\bibitem [{\citenamefont {Herschbach}(1982)}]{MIT}%
  \BibitemOpen
  \bibfield  {author} {\bibinfo {author} {\bibfnamefont {D.}~\bibnamefont
  {Herschbach}},\ }\bibfield  {title} {\bibinfo {title} {{The Van Vleck
  transformation in perturbation theory}},\ }\href
  {https://ocw.mit.edu/courses/5-80-small-molecule-spectroscopy-and-dynamics-fall-2008/resources/vanvleck_1982/}
  {\bibfield  {journal} {\bibinfo  {journal} {MIT Lecture Notes}\ } (\bibinfo
  {year} {1982})}\BibitemShut {NoStop}%
\bibitem [{\citenamefont {Cazalilla}(2003)}]{Cazalilla}%
  \BibitemOpen
  \bibfield  {author} {\bibinfo {author} {\bibfnamefont {M.~A.}\ \bibnamefont
  {Cazalilla}},\ }\bibfield  {title} {\bibinfo {title} {{One-dimensional
  optical lattices and impenetrable bosons}},\ }\href
  {https://doi.org/10.1103/PhysRevA.67.053606} {\bibfield  {journal} {\bibinfo
  {journal} {Phys. Rev. A}\ }\textbf {\bibinfo {volume} {67}},\ \bibinfo
  {pages} {053606} (\bibinfo {year} {2003})}\BibitemShut {NoStop}%
\bibitem [{\citenamefont {Basak}\ and\ \citenamefont
  {Pu}(2021)}]{basak2021strongly}%
  \BibitemOpen
  \bibfield  {author} {\bibinfo {author} {\bibfnamefont {S.}~\bibnamefont
  {Basak}}\ and\ \bibinfo {author} {\bibfnamefont {H.}~\bibnamefont {Pu}},\
  }\bibfield  {title} {\bibinfo {title} {{Strongly interacting two-component
  coupled Bose gas in optical lattices}},\ }\href
  {https://journals.aps.org/pra/abstract/10.1103/PhysRevA.104.053326}
  {\bibfield  {journal} {\bibinfo  {journal} {Phys. Rev. A}\ }\textbf {\bibinfo
  {volume} {104}},\ \bibinfo {pages} {053326} (\bibinfo {year}
  {2021})}\BibitemShut {NoStop}%
\bibitem [{\citenamefont {Zhang}\ and\ \citenamefont {Dong}(2010)}]{Zhang2010}%
  \BibitemOpen
  \bibfield  {author} {\bibinfo {author} {\bibfnamefont {J.~M.}\ \bibnamefont
  {Zhang}}\ and\ \bibinfo {author} {\bibfnamefont {R.~X.}\ \bibnamefont
  {Dong}},\ }\bibfield  {title} {\bibinfo {title} {{Exact diagonalization: the
  Bose{\textendash}Hubbard model as an example}},\ }\href
  {https://doi.org/10.1088/0143-0807/31/3/016} {\bibfield  {journal} {\bibinfo
  {journal} {European Journal of Physics}\ }\textbf {\bibinfo {volume} {31}},\
  \bibinfo {pages} {591} (\bibinfo {year} {2010})}\BibitemShut {NoStop}%
\bibitem [{\citenamefont {Oganesyan}\ and\ \citenamefont
  {Huse}(2007)}]{Oganesyan}%
  \BibitemOpen
  \bibfield  {author} {\bibinfo {author} {\bibfnamefont {V.}~\bibnamefont
  {Oganesyan}}\ and\ \bibinfo {author} {\bibfnamefont {D.~A.}\ \bibnamefont
  {Huse}},\ }\bibfield  {title} {\bibinfo {title} {{Localization of interacting
  fermions at high temperature}},\ }\href
  {https://doi.org/10.1103/PhysRevB.75.155111} {\bibfield  {journal} {\bibinfo
  {journal} {Phys. Rev. B}\ }\textbf {\bibinfo {volume} {75}},\ \bibinfo
  {pages} {155111} (\bibinfo {year} {2007})}\BibitemShut {NoStop}%
\bibitem [{\citenamefont {Atas}\ \emph {et~al.}(2013)\citenamefont {Atas},
  \citenamefont {Bogomolny}, \citenamefont {Giraud},\ and\ \citenamefont
  {Roux}}]{Atas}%
  \BibitemOpen
  \bibfield  {author} {\bibinfo {author} {\bibfnamefont {Y.~Y.}\ \bibnamefont
  {Atas}}, \bibinfo {author} {\bibfnamefont {E.}~\bibnamefont {Bogomolny}},
  \bibinfo {author} {\bibfnamefont {O.}~\bibnamefont {Giraud}},\ and\ \bibinfo
  {author} {\bibfnamefont {G.}~\bibnamefont {Roux}},\ }\bibfield  {title}
  {\bibinfo {title} {{Distribution of the Ratio of Consecutive Level Spacings
  in Random Matrix Ensembles}},\ }\href
  {https://doi.org/10.1103/PhysRevLett.110.084101} {\bibfield  {journal}
  {\bibinfo  {journal} {Phys. Rev. Lett.}\ }\textbf {\bibinfo {volume} {110}},\
  \bibinfo {pages} {084101} (\bibinfo {year} {2013})}\BibitemShut {NoStop}%
\bibitem [{\citenamefont {Chen}\ \emph {et~al.}(2024)\citenamefont {Chen},
  \citenamefont {Chen},\ and\ \citenamefont {Wang}}]{chen2024eigenstate}%
  \BibitemOpen
  \bibfield  {author} {\bibinfo {author} {\bibfnamefont {J.}~\bibnamefont
  {Chen}}, \bibinfo {author} {\bibfnamefont {C.}~\bibnamefont {Chen}},\ and\
  \bibinfo {author} {\bibfnamefont {X.}~\bibnamefont {Wang}},\ }\bibfield
  {title} {\bibinfo {title} {{Eigenstate properties of the disordered
  Bose--Hubbard chain}},\ }\href
  {https://link.springer.com/article/10.1007/s11467-023-1384-1} {\bibfield
  {journal} {\bibinfo  {journal} {Front. Phys.}\ }\textbf {\bibinfo {volume}
  {19}},\ \bibinfo {pages} {43207} (\bibinfo {year} {2024})}\BibitemShut
  {NoStop}%
\bibitem [{\citenamefont {Serbyn}\ \emph {et~al.}(2013)\citenamefont {Serbyn},
  \citenamefont {Papi\'c},\ and\ \citenamefont {Abanin}}]{Serbyn}%
  \BibitemOpen
  \bibfield  {author} {\bibinfo {author} {\bibfnamefont {M.}~\bibnamefont
  {Serbyn}}, \bibinfo {author} {\bibfnamefont {Z.}~\bibnamefont {Papi\'c}},\
  and\ \bibinfo {author} {\bibfnamefont {D.~A.}\ \bibnamefont {Abanin}},\
  }\bibfield  {title} {\bibinfo {title} {{Local Conservation Laws and the
  Structure of the Many-Body Localized States}},\ }\href
  {https://doi.org/10.1103/PhysRevLett.111.127201} {\bibfield  {journal}
  {\bibinfo  {journal} {Phys. Rev. Lett.}\ }\textbf {\bibinfo {volume} {111}},\
  \bibinfo {pages} {127201} (\bibinfo {year} {2013})}\BibitemShut {NoStop}%
\bibitem [{\citenamefont {Huse}\ \emph {et~al.}(2014)\citenamefont {Huse},
  \citenamefont {Nandkishore},\ and\ \citenamefont {Oganesyan}}]{HuseLIOM}%
  \BibitemOpen
  \bibfield  {author} {\bibinfo {author} {\bibfnamefont {D.~A.}\ \bibnamefont
  {Huse}}, \bibinfo {author} {\bibfnamefont {R.}~\bibnamefont {Nandkishore}},\
  and\ \bibinfo {author} {\bibfnamefont {V.}~\bibnamefont {Oganesyan}},\
  }\bibfield  {title} {\bibinfo {title} {{Phenomenology of fully
  many-body-localized systems}},\ }\href
  {https://doi.org/10.1103/PhysRevB.90.174202} {\bibfield  {journal} {\bibinfo
  {journal} {Phys. Rev. B}\ }\textbf {\bibinfo {volume} {90}},\ \bibinfo
  {pages} {174202} (\bibinfo {year} {2014})}\BibitemShut {NoStop}%
\bibitem [{\citenamefont {Chandran}\ \emph {et~al.}(2015)\citenamefont
  {Chandran}, \citenamefont {Kim}, \citenamefont {Vidal},\ and\ \citenamefont
  {Abanin}}]{Anushya}%
  \BibitemOpen
  \bibfield  {author} {\bibinfo {author} {\bibfnamefont {A.}~\bibnamefont
  {Chandran}}, \bibinfo {author} {\bibfnamefont {I.~H.}\ \bibnamefont {Kim}},
  \bibinfo {author} {\bibfnamefont {G.}~\bibnamefont {Vidal}},\ and\ \bibinfo
  {author} {\bibfnamefont {D.~A.}\ \bibnamefont {Abanin}},\ }\bibfield  {title}
  {\bibinfo {title} {{Constructing local integrals of motion in the many-body
  localized phase}},\ }\href {https://doi.org/10.1103/PhysRevB.91.085425}
  {\bibfield  {journal} {\bibinfo  {journal} {Phys. Rev. B}\ }\textbf {\bibinfo
  {volume} {91}},\ \bibinfo {pages} {085425} (\bibinfo {year}
  {2015})}\BibitemShut {NoStop}%
\bibitem [{\citenamefont {Geraedts}\ \emph {et~al.}(2017)\citenamefont
  {Geraedts}, \citenamefont {Bhatt},\ and\ \citenamefont
  {Nandkishore}}]{geraedts2017emergent}%
  \BibitemOpen
  \bibfield  {author} {\bibinfo {author} {\bibfnamefont {S.~D.}\ \bibnamefont
  {Geraedts}}, \bibinfo {author} {\bibfnamefont {R.~N.}\ \bibnamefont
  {Bhatt}},\ and\ \bibinfo {author} {\bibfnamefont {R.}~\bibnamefont
  {Nandkishore}},\ }\bibfield  {title} {\bibinfo {title} {{Emergent local
  integrals of motion without a complete set of localized eigenstates}},\
  }\href {https://journals.aps.org/prb/abstract/10.1103/PhysRevB.95.064204}
  {\bibfield  {journal} {\bibinfo  {journal} {Phys. Rev. B}\ }\textbf {\bibinfo
  {volume} {95}},\ \bibinfo {pages} {064204} (\bibinfo {year}
  {2017})}\BibitemShut {NoStop}%
\bibitem [{\citenamefont {Sala}\ \emph {et~al.}(2020)\citenamefont {Sala},
  \citenamefont {Rakovszky}, \citenamefont {Verresen}, \citenamefont {Knap},\
  and\ \citenamefont {Pollmann}}]{SalaPollmann}%
  \BibitemOpen
  \bibfield  {author} {\bibinfo {author} {\bibfnamefont {P.}~\bibnamefont
  {Sala}}, \bibinfo {author} {\bibfnamefont {T.}~\bibnamefont {Rakovszky}},
  \bibinfo {author} {\bibfnamefont {R.}~\bibnamefont {Verresen}}, \bibinfo
  {author} {\bibfnamefont {M.}~\bibnamefont {Knap}},\ and\ \bibinfo {author}
  {\bibfnamefont {F.}~\bibnamefont {Pollmann}},\ }\bibfield  {title} {\bibinfo
  {title} {{Ergodicity Breaking Arising from Hilbert Space Fragmentation in
  Dipole-Conserving Hamiltonians}},\ }\href
  {https://doi.org/10.1103/PhysRevX.10.011047} {\bibfield  {journal} {\bibinfo
  {journal} {Phys. Rev. X}\ }\textbf {\bibinfo {volume} {10}},\ \bibinfo
  {pages} {011047} (\bibinfo {year} {2020})}\BibitemShut {NoStop}%
\bibitem [{\citenamefont {Huber}\ and\ \citenamefont
  {Lindner}(2011)}]{huber2011topological}%
  \BibitemOpen
  \bibfield  {author} {\bibinfo {author} {\bibfnamefont {S.~D.}\ \bibnamefont
  {Huber}}\ and\ \bibinfo {author} {\bibfnamefont {N.~H.}\ \bibnamefont
  {Lindner}},\ }\bibfield  {title} {\bibinfo {title} {{Topological transitions
  for lattice bosons in a magnetic field}},\ }\href
  {https://www.pnas.org/doi/abs/10.1073/pnas.1110813108} {\bibfield  {journal}
  {\bibinfo  {journal} {Proc. Natl. Acad. Sci.}\ }\textbf {\bibinfo {volume}
  {108}},\ \bibinfo {pages} {19925} (\bibinfo {year} {2011})}\BibitemShut
  {NoStop}%
\bibitem [{\citenamefont {Lacroix}\ \emph {et~al.}(2011)\citenamefont
  {Lacroix}, \citenamefont {Mendels},\ and\ \citenamefont
  {Mila}}]{lacroix2011introduction}%
  \BibitemOpen
  \bibfield  {author} {\bibinfo {author} {\bibfnamefont {C.}~\bibnamefont
  {Lacroix}}, \bibinfo {author} {\bibfnamefont {P.}~\bibnamefont {Mendels}},\
  and\ \bibinfo {author} {\bibfnamefont {F.}~\bibnamefont {Mila}},\ }\href@noop
  {} {\emph {\bibinfo {title} {{Introduction to frustrated magnetism:
  materials, experiments, theory}}}},\ Vol.\ \bibinfo {volume} {164, p. 537}\
  (\bibinfo  {publisher} {Springer Science \& Business Media},\ \bibinfo {year}
  {2011})\BibitemShut {NoStop}%
\bibitem [{\citenamefont {Molignini}(2025)}]{Molignini_arxiv}%
  \BibitemOpen
  \bibfield  {author} {\bibinfo {author} {\bibfnamefont {P.}~\bibnamefont
  {Molignini}},\ }\bibfield  {title} {\bibinfo {title} {{Stability of
  quasicrystalline ultracold fermions to dipolar interactions}},\ }\href
  {https://doi.org/10.1103/szdc-61nl} {\bibfield  {journal} {\bibinfo
  {journal} {Phys. Rev. Res.}\ }\textbf {\bibinfo {volume} {7}},\ \bibinfo
  {pages} {L032026} (\bibinfo {year} {2025})}\BibitemShut {NoStop}%
\bibitem [{\citenamefont {Carleo}\ \emph {et~al.}(2012)\citenamefont {Carleo},
  \citenamefont {Becca}, \citenamefont {Schir\'o},\ and\ \citenamefont
  {Fabrizio}}]{Carleo}%
  \BibitemOpen
  \bibfield  {author} {\bibinfo {author} {\bibfnamefont {G.}~\bibnamefont
  {Carleo}}, \bibinfo {author} {\bibfnamefont {F.}~\bibnamefont {Becca}},
  \bibinfo {author} {\bibfnamefont {M.}~\bibnamefont {Schir\'o}},\ and\
  \bibinfo {author} {\bibfnamefont {M.}~\bibnamefont {Fabrizio}},\ }\bibfield
  {title} {\bibinfo {title} {{Localization and Glassy Dynamics Of Many-Body
  Quantum Systems}},\ }\href {https://doi.org/10.1038/srep00243} {\bibfield
  {journal} {\bibinfo  {journal} {Sci. Rep.}\ }\textbf {\bibinfo {volume}
  {2}},\ \bibinfo {pages} {243} (\bibinfo {year} {2012})}\BibitemShut {NoStop}%
\bibitem [{\citenamefont {Honda}\ \emph {et~al.}(2025)\citenamefont {Honda},
  \citenamefont {Takasu}, \citenamefont {Goto}, \citenamefont {Kazuta},
  \citenamefont {Kunimi}, \citenamefont {Danshita},\ and\ \citenamefont
  {Takahashi}}]{Honda}%
  \BibitemOpen
  \bibfield  {author} {\bibinfo {author} {\bibfnamefont {K.}~\bibnamefont
  {Honda}}, \bibinfo {author} {\bibfnamefont {Y.}~\bibnamefont {Takasu}},
  \bibinfo {author} {\bibfnamefont {S.}~\bibnamefont {Goto}}, \bibinfo {author}
  {\bibfnamefont {H.}~\bibnamefont {Kazuta}}, \bibinfo {author} {\bibfnamefont
  {M.}~\bibnamefont {Kunimi}}, \bibinfo {author} {\bibfnamefont
  {I.}~\bibnamefont {Danshita}},\ and\ \bibinfo {author} {\bibfnamefont
  {Y.}~\bibnamefont {Takahashi}},\ }\bibfield  {title} {\bibinfo {title}
  {{Observation of slow relaxation due to Hilbert space fragmentation in
  strongly interacting Bose-Hubbard chains}},\ }\href
  {https://doi.org/10.1126/sciadv.adv3255} {\bibfield  {journal} {\bibinfo
  {journal} {Science Advances}\ }\textbf {\bibinfo {volume} {11}},\ \bibinfo
  {pages} {eadv3255} (\bibinfo {year} {2025})}\BibitemShut {NoStop}%
\bibitem [{\citenamefont {Kollath}\ \emph {et~al.}(2007)\citenamefont
  {Kollath}, \citenamefont {L\"auchli},\ and\ \citenamefont
  {Altman}}]{Kollath}%
  \BibitemOpen
  \bibfield  {author} {\bibinfo {author} {\bibfnamefont {C.}~\bibnamefont
  {Kollath}}, \bibinfo {author} {\bibfnamefont {A.~M.}\ \bibnamefont
  {L\"auchli}},\ and\ \bibinfo {author} {\bibfnamefont {E.}~\bibnamefont
  {Altman}},\ }\bibfield  {title} {\bibinfo {title} {{Quench Dynamics and
  Nonequilibrium Phase Diagram of the Bose-Hubbard Model}},\ }\href
  {https://doi.org/10.1103/PhysRevLett.98.180601} {\bibfield  {journal}
  {\bibinfo  {journal} {Phys. Rev. Lett.}\ }\textbf {\bibinfo {volume} {98}},\
  \bibinfo {pages} {180601} (\bibinfo {year} {2007})}\BibitemShut {NoStop}%
\bibitem [{\citenamefont {Chen}\ \emph {et~al.}(2025)\citenamefont {Chen},
  \citenamefont {Chen},\ and\ \citenamefont {Wang}}]{jiechen2023}%
  \BibitemOpen
  \bibfield  {author} {\bibinfo {author} {\bibfnamefont {J.}~\bibnamefont
  {Chen}}, \bibinfo {author} {\bibfnamefont {C.}~\bibnamefont {Chen}},\ and\
  \bibinfo {author} {\bibfnamefont {X.}~\bibnamefont {Wang}},\ }\bibfield
  {title} {\bibinfo {title} {{Symmetry- and energy-resolved entanglement
  dynamics in a disordered Bose-Hubbard model}},\ }\href
  {https://doi.org/10.1103/88sq-cm27} {\bibfield  {journal} {\bibinfo
  {journal} {Phys. Rev. B}\ }\textbf {\bibinfo {volume} {112}},\ \bibinfo
  {pages} {094201} (\bibinfo {year} {2025})}\BibitemShut {NoStop}%
\bibitem [{\citenamefont {Sarkar}\ and\ \citenamefont {Sen}(2022)}]{Sarkar}%
  \BibitemOpen
  \bibfield  {author} {\bibinfo {author} {\bibfnamefont {S.}~\bibnamefont
  {Sarkar}}\ and\ \bibinfo {author} {\bibfnamefont {U.}~\bibnamefont {Sen}},\
  }\bibfield  {title} {\bibinfo {title} {Glassy disorder-induced effects in
  noisy dynamics of bose–hubbard and fermi–hubbard systems},\ }\href
  {https://doi.org/10.1088/1361-6455/ac8e3b} {\bibfield  {journal} {\bibinfo
  {journal} {Journal of Physics B: Atomic, Molecular and Optical Physics}\
  }\textbf {\bibinfo {volume} {55}},\ \bibinfo {pages} {205502} (\bibinfo
  {year} {2022})}\BibitemShut {NoStop}%
\bibitem [{\citenamefont {Sierant}\ \emph
  {et~al.}(2017{\natexlab{a}})\citenamefont {Sierant}, \citenamefont
  {Delande},\ and\ \citenamefont {Zakrzewski}}]{Sierant2017}%
  \BibitemOpen
  \bibfield  {author} {\bibinfo {author} {\bibfnamefont {P.}~\bibnamefont
  {Sierant}}, \bibinfo {author} {\bibfnamefont {D.}~\bibnamefont {Delande}},\
  and\ \bibinfo {author} {\bibfnamefont {J.}~\bibnamefont {Zakrzewski}},\
  }\bibfield  {title} {\bibinfo {title} {{Many-body localization due to random
  interactions}},\ }\href {https://doi.org/10.1103/PhysRevA.95.021601}
  {\bibfield  {journal} {\bibinfo  {journal} {Phys. Rev. A}\ }\textbf {\bibinfo
  {volume} {95}},\ \bibinfo {pages} {021601} (\bibinfo {year}
  {2017}{\natexlab{a}})}\BibitemShut {NoStop}%
\bibitem [{\citenamefont {Sierant}\ \emph
  {et~al.}(2017{\natexlab{b}})\citenamefont {Sierant}, \citenamefont
  {Delande},\ and\ \citenamefont {Zakrzewski}}]{Sierant2017Poland}%
  \BibitemOpen
  \bibfield  {author} {\bibinfo {author} {\bibfnamefont {P.}~\bibnamefont
  {Sierant}}, \bibinfo {author} {\bibfnamefont {D.}~\bibnamefont {Delande}},\
  and\ \bibinfo {author} {\bibfnamefont {J.}~\bibnamefont {Zakrzewski}},\
  }\bibfield  {title} {\bibinfo {title} {Many-body localization for randomly
  interacting bosons},\ }\href {https://doi.org/10.12693/APhysPolA.132.1707}
  {\bibfield  {journal} {\bibinfo  {journal} {Acta Physica Polonica A}\
  }\textbf {\bibinfo {volume} {132}},\ \bibinfo {pages} {1707} (\bibinfo {year}
  {2017}{\natexlab{b}})}\BibitemShut {NoStop}%
\bibitem [{\citenamefont {Molignini}\ \emph {et~al.}(2023)\citenamefont
  {Molignini}, \citenamefont {Arandes},\ and\ \citenamefont
  {Bergholtz}}]{Molignini_skin}%
  \BibitemOpen
  \bibfield  {author} {\bibinfo {author} {\bibfnamefont {P.}~\bibnamefont
  {Molignini}}, \bibinfo {author} {\bibfnamefont {O.}~\bibnamefont {Arandes}},\
  and\ \bibinfo {author} {\bibfnamefont {E.~J.}\ \bibnamefont {Bergholtz}},\
  }\bibfield  {title} {\bibinfo {title} {{Anomalous skin effects in disordered
  systems with a single non-Hermitian impurity}},\ }\href
  {https://doi.org/10.1103/PhysRevResearch.5.033058} {\bibfield  {journal}
  {\bibinfo  {journal} {Phys. Rev. Res.}\ }\textbf {\bibinfo {volume} {5}},\
  \bibinfo {pages} {033058} (\bibinfo {year} {2023})}\BibitemShut {NoStop}%
\end{thebibliography}%

\end{document}